\documentclass{article}
\usepackage{graphicx} 
\usepackage{siunitx}
\usepackage{listings}
\usepackage{hyperref}
\usepackage{nomencl} 
\makenomenclature
\usepackage{amsmath}
\usepackage{algorithm2e}
\usepackage{bm}
\usepackage{booktabs}
\usepackage[letterpaper, total={6in, 9in}]{geometry}

\title{One-Cell Inversion for Solving Higher-Order Time-Dependent Radiation Transport on GPUs
\footnote{This is an Accepted Manuscript of an article accepted to be published by Taylor \& Francis in Nuclear Science and Engineering.}} 

\author{
  Joanna Piper Morgan$^{1,2}$\footnote{Contact: morgajoa@oregonstate.edu, joannapipermorgan@gmail.com}
  \and
  Ilham Variansyah$^{1,3}$
  \and
  Todd S. Palmer$^{1,3}$
  \and
  Kyle E. Niemeyer$^{1,2}$
}

\date{%
    \small{
    $^1$Center for Exascale Monte Carlo Neutron Transport (CEMeNT)\footnote{https://cement-psaap.github.io/}\\
    $^2$School of Mechanical Industrial and Manufacturing Engineering, Oregon State University, Corvallis, OR, 97331, USA\\%
    $^3$School of Nuclear Science and Engineering, Oregon State University, Corvallis, OR, 97331, USA%
    }
}

\begin{document}

\maketitle

\begin{abstract}
    To find deterministic solutions to the transient discrete-ordinates neutron-transport equation, source iterations (SI) are typically used to lag the scattering (and fission) source terms from subsequent iterations.
    For Cartesian geometries in one dimension, SI is parallel over the number of angles but not spatial cells; this is a disadvantage for many-core compute architectures like graphics processing units.
    One-cell inversion (OCI) is a class of alternative iterative methods that allow space-parallel radiation transport on heterogeneous compute architectures.
    For OCI, previous studies have shown that, in steady-state computations, spectral radius tends to unity when cells are optically thin regardless of the scattering ratio.
    In this work, we analyze how the convergence rate of an OCI scheme behaves when used for time-dependent neutron transport computations.
    We derive a second-order space-time discretization method from the simple corner balance and multiple balance time discretization schemes and show via Fourier analysis that it is unconditionally stable through time.
    Then, we derive and numerically solve the Fourier systems for both OCI and SI splittings of our discretization, showing that small mean-free times improve the spectral radius of OCI more than SI, and that spectral radius for OCI tends to zero as mean free time gets smaller.
    We extend both solvers to be energy dependent (using the multi-group assumption) and implement on an AMD MI250X using vendor-supplied batched LAPACK solvers.
    Smaller time steps improve the relative performance of OCI over SI, and, even when OCI requires more iterations to converge a problem, those iterations can be done much faster on a GPU.
    This leads to OCI performing better overall than SI on GPUs.
\end{abstract}

\section{Introduction}

Simulating transient particle transport is often required when computing the solution to a number of multi-physics problems, including burst criticality experiments, fission reactor accidents, and other highly dynamic systems.
Finding deterministic solutions to the transient neutron transport equation requires some method of treating the contribution of scattering described by an integral.
This is either done by taking moments of the neutron transport equation and making a closure assumption, or by using quadrature to discretize the integral over angle.
The latter is called the method of discrete ordinates (or S$_N$ method, where $N$ is the number of angles in a one-dimensional quadrature set) that forms a coupled set of simultaneous PDEs, with one for every direction in a given quadrature set.
The contribution to the scattering source can then be computed using a sum over angles of weights times quantities of interest.
Typically, iterative schemes from operator splitting are used to treat the scattering (and fission) source terms that arise in this coupled set of partial differential equations \cite{lewis1984computational}.

Source iteration (SI), often accompanied by preconditioners or synthetic accelerators, is a common iteration approach: the contribution to the solution from the scattering source lags, while the angular flux is solved in every ordinate direction via ``sweeps'' through the spatial domain~\cite{adams_subcell_1997}.
SI sweeps in Cartesian geometries readily parallelize over the number of angles.
While any parallelization improves performance, a scheme that is embarrassingly parallel over the dimension with the greatest number of degrees of freedom---space---would be advantageous, especially on vectorized hardware \cite{rosa_cellwise_2013, hoagland_hybrid_2021}.
In slab geometry, SI sweeps can be parallelized in angle and energy groups (via Jacobi iteration), but are serial in space as information about the angular flux incident on edges of each cell is required before the computation can proceed.

In higher spatial dimensions, many S$_N$ production codes that implement SI use a wavefront marching parallel algorithm known as a Koch--Baker--Alcouffe scheme \cite{baker_sn_2017}, also called ``full parallel sweeps.''
This algorithm begins a sweep in a spatial location where all cell dependencies are known from boundary information (e.g., a corner).
From there, on a hypothetical orthogonal 2D spatial grid the two nearest neighbor cells are computed independently in parallel; the next step would be across four cells.
This diagonally expanding wavefront continues to march and can compute quantities of interest in parallel for as many cells that lie on the diagonal sweep step.
These sweeps are done on structured or unstructured finite element or finite volume spatial discretizations with backward Euler or Crank--Nicholson time stepping iterations.
Source iteration is often solved with preconditioned fixed-point (Richardson) or Krylov sub-space methods (e.g., GMRES) \cite{adams_fast_2002}.

An alternative to SI is one-cell inversion (OCI), a class of operator splitting that computes all angular fluxes in all ordinates within a cell in a single linear algebra solve, assuming that the angular fluxes incident on the surfaces of the cell are known from a previous iteration \cite{kang2000oci}.
OCI methods allow parallelizing over the number of cells, as each cell is solved independently in parallel.
OCI iterations can take the form of a cell-wise block-Jacobi, cell-wise block-Gauss--Seidel, or cell-wise red-black iteration depending on the order in which cells are inverted \cite{man1994parallel}.
Like SI, OCI iterations can be fixed-point (Richardson) or non-stationary schemes like GMRES \cite{kylov2004warsa}, with or without preconditioners (including diffusion synthetic acceleration \cite{kang2000oci}) on structured and unstructured meshes.
Parallel block Jacobi and parallel block Gauss--Seidel iterations may also be used for domain decomposition with transport sweeps within subdomains \cite{qiao_improved_2021}.
In fact, OCI methods can be thought of as a cellular decomposed version of these schemes.

\cite{rosa_cellwise_2013} previously studied cell-wise block Jacobi and cell-wise block Gauss--Seidel as a potentially superior iterative scheme over SI preconditioned with diffusion synthetic acceleration on vectorized architectures.
They hypothesized that OCI schemes might outperform an SI preconditioned with diffusion synthetic acceleration and using full-parallel sweeps in terms of wall-clock runtime, because of OCI's parallelism over the dominant domain (space), the ability to take advantage of vendor-supplied LAPACK type libraries, high arithmetic-intensity operations present in an OCI algorithm, and superior spectral properties in the thick limit.
Rosa et al.\ conducted Fourier analysis for and implemented OCI in a 2D, multi-group, steady-state code using bilinear discontinuous finite elements to discretize space.
They paired this with either a cell-wise block Jacobi and cell-wise block Gauss--Seidel iteration algorithm.
The study was conducted on the (then) state-of-the-art RoadRunner supercomputer and took advantage of its 64-bit PowerXCell vectorized accelerator.
However, the acceleration per iteration in the block Gauss--Seidel OCI implementation did not make up for the degradation of convergence that OCI methods incur in the thin limit.

\begin{figure}
    \centering
    \includegraphics[width=\textwidth]{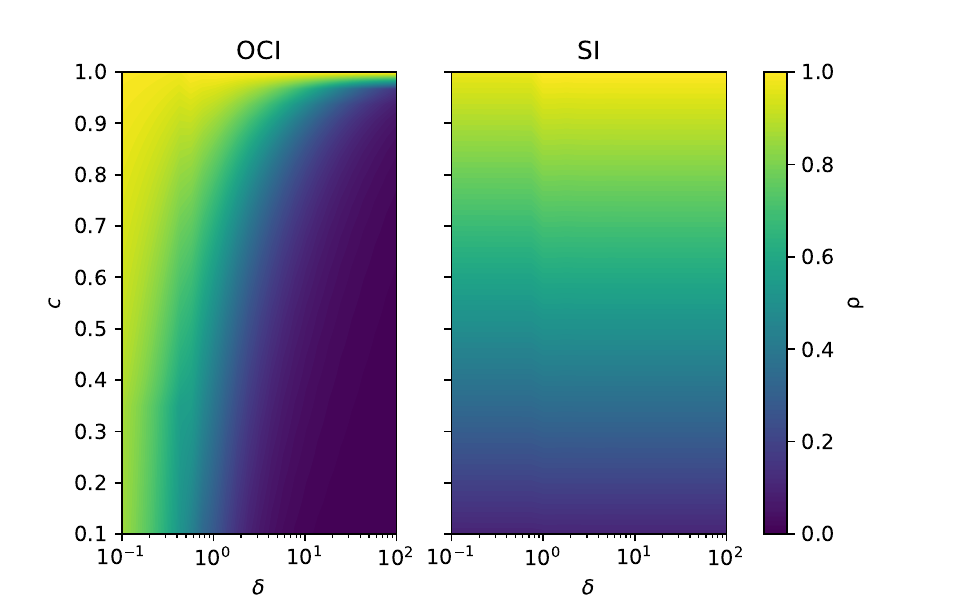}
    \caption{Spectral radii (${\rho}$) of steady-state OCI (left) and SI (right), where $c$ is the scattering ratio and ${\delta}$ is the cellular optical thickness in mean free paths from Fourier analysis in S$_8$.}
    \label{fig:ss-sepcrad}
\end{figure}

OCI can require more iterations to converge to a solution for some problems, since no information exchanges between cells within an iteration.
Specifically, as cellular optical thickness decreases, OCI's relative performance degrades.
Spectral radius ($\rho$) of OCI tends to unity in the thin cell limit---regardless of the scattering ratio---due to the algorithm decoupling cells from one another (i.e., asynchronicity) \cite{rosa_cellwise_2013, hoagland_hybrid_2021, man1994parallel}. 
Figure~\ref{fig:ss-sepcrad} illustrates this behavior, showing the spectral radii of the two iteration schemes as a function of cellular optical thickness, $\delta$ (in mean free paths), and the scattering ratio, $c$.
We compute these values using Fourier analysis of an infinite medium slab problem using S$_{8}$ angular quadrature for block Jacobi OCI and unpreconditioned SI iterative schemes\footnote{Gauss--Legendre quadrature is used in all presented work.}.
The spectral radius of SI depends strongly on the scattering ratio but is independent of $\delta$ for the homogeneous infinite-medium problem. 
Compared to SI, OCI rapidly converges in thicker cells, even in highly scattering problems except for scattering ratios closest to one.

Others have explored OCI as an acceleration scheme for SI \cite{ hoagland_hybrid_2021}, a component of a multi-grid solver \cite{man1995multigrid1, man1996multigrid2}, and a solution to the integral transport matrix method \cite{raffi2108pidotscom}.
However, previous investigations of OCI are limited to steady-state computations.

When employing implicit time-differencing schemes (e.g., Crank--Nicholson, backward Euler), each time step involves the solution of a steady-state transport problem with an effective total cross section that includes a time absorption term, proportional to $1/(v \Delta t)$, where $\Delta t$ is time step size and $v$ is radiation speed.
Returning to Fig.~\ref{fig:ss-sepcrad}, the macroscopic total cross section ($\Sigma$) influences both the optical thickness of the cell ($\delta$) and the scattering ratio ($c$), so increasing or decreasing $\Sigma$ will impact convergence behavior.
Spectral radius for both iterative methods decreases as the scattering ratio decreases, but \textit{the spectral radius of OCI also decreases with increasing optical thickness}, which in turn depends on $\Sigma$.
When solving optically thin and highly scattering problems, small increases to $\Sigma$ (and for time-dependent problems, decreases in $\Delta t$ or $v$) may drastically improve the relative performance of OCI compared to SI.
This hypothesis motivates our work, along with evaluating cell-wise algorithms on modern GPU accelerators and exploring higher-order space-time discretization schemes.

We previously derived a second-order space (simple corner balance), and time discretization (multiple balance) scheme for block Jacobi OCI (which we will call simply OCI in the remainder of this work) \cite{morgan2023oci}.
We previously showed that when there are more spatial degrees of freedom than in angle, a GPU implementation of OCI will outperform a similarly implemented version of SI in wall clock runtime, in all but the highest scattering problems, for quadrature orders between \num{16} and \num{64} for mono-energetic 1D problems. 

Some derivations from our previous work are included here (Section~\ref{sec:methods-derv}), because we extend it with a Fourier analysis of the discretization scheme through time to ensure it remains unconditionally stable.
We also perform a Fourier analysis on a single time step of OCI and SI to study convergence behaviors in various limits under transient conditions.
Furthermore we extend our derivations to multi-group problems and implement both OCI and SI on an AMD MI250X GPU using vendor-supplied libraries to confirm Fourier results and analyze performance.

\section{Methods}

In this section, we derive the discretized equations for the initial and boundary value problem and describe the OCI iteration.
We have chosen to implement robust second-order discretization methods: simple corner balance  \cite{adams_subcell_1997} in space and multiple balance \cite{variansyah_robust_2021} in time.
By coupling these higher accuracy schemes with an efficient iterative method, we hope to optimize the ratio of compute work to communication work to better suit the numerical method for GPUs.
To confirm that multiple balance time discretization remains unconditionally stable with simple corner balance, we conduct Fourier analysis for a non-scattering model problem.
Then, we derive the Fourier system for a single time step of simple corner balance + multiple balance discretization using both OCI and SI operator splitting to study the convergence rate.
Finally, we present systems for multi-group transport.

\subsection{Derivation of space and time discretization for one-cell inversion}
\label{sec:methods-derv}
We begin with the time-dependent, isotropic scattering slab geometry, S$_N$ transport equations with an isotropic source.
\begin{multline}
    \label{eq:sn_nte}
    \frac{1}{v} \frac{\partial \psi_{m}(x,t)}{\partial t} + \mu_m \frac{\partial \psi_{m}(x,t)}{\partial x} + \Sigma(x) \psi_{m}(x,t) 
     = \frac{1}{2} \left( \Sigma_{s}(x) \sum\limits_{n=1}^N w_n \psi_{n}(x,t) + Q(x,t) \right) \;, \\
    \qquad \qquad m=1, \ldots, N \;, \qquad t > 0 \;, \qquad x \in [0,X]
\end{multline}
where $\psi$ is the angular flux, $t$ is time, $x$ is location, $v$ is speed, $\Sigma$ is the macroscopic total cross-section, $\Sigma_s$ is the macroscopic scattering cross-section, $w_m$ is angular quadrature weight, $\mu_m$ is the angular quadrature ordinate, $m$ is the quadrature index, $N$ is the quadrature order, and $Q$ is the isotropic material source.
The initial and boundary conditions are prescribed angular flux distributions:
\begin{equation*}
    \psi_{m}(x,0) = \psi_{init,m}(x), \qquad m=1 \ldots N \;,
\end{equation*}
\begin{equation*}
    \psi_{m}(0,t) = \psi_{inc,m}^+(t), \qquad \mu_m >0 \;,
\end{equation*}
\begin{equation*}
    \psi_{m}(X,t) = \psi_{inc,m}^-(t), \qquad \mu_m <0 \;.
\end{equation*}

\begin{figure}
    \centering
    \includegraphics[width=\textwidth]{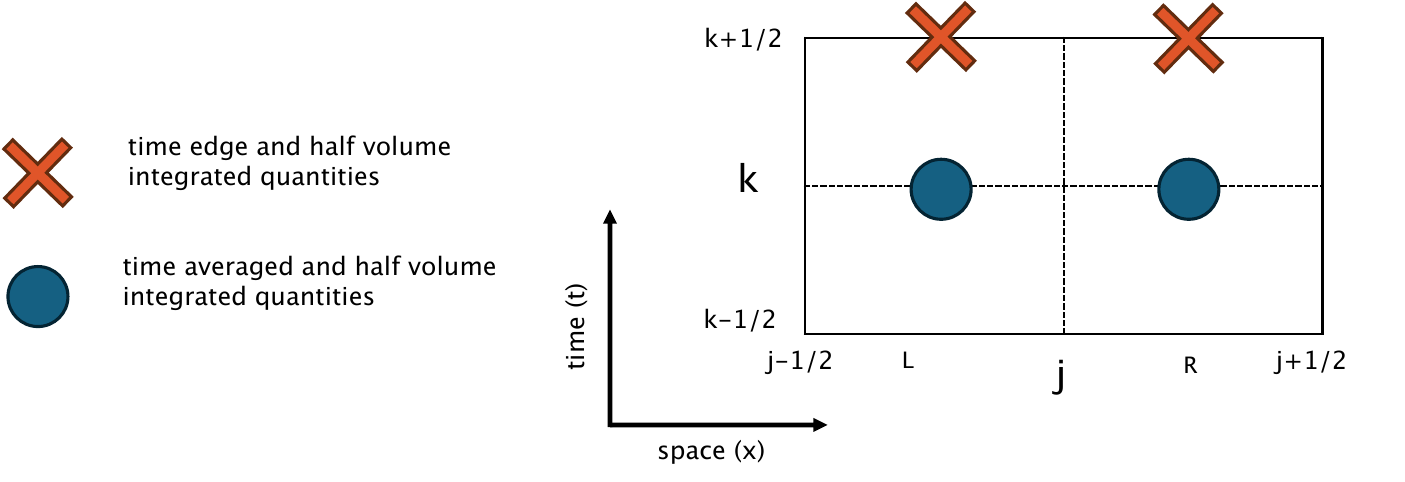}
    \caption{Discretization stencil for simple corner balance, multiple balance time discretization}
    \label{fig:stencil}
\end{figure}

We discretize these equations in time using multiple balance~\cite{variansyah_robust_2021}, which solves two coupled sets of equations. 
First is a backward Euler step (transport equation integrated over a time step):
\begin{subequations}
\begin{multline}
\frac{1}{v} \left( \frac{\psi_{m,k+1/2}(x) - \psi_{m,k-1/2}(x)}{\Delta t} \right) + \mu_m \frac{\partial \psi_{m,k}(x)}{\partial x} + \Sigma(x) \psi_{m,k}(x) \\
= \frac{1}{2} \left(  \Sigma_{s}(x) \sum\limits_{n=1}^N w_n \psi_{n,k}(x) + Q_{k}(x) \right) \;,
\end{multline}
and the second is a balance like auxiliary equation from the multiple balance principle:
\begin{multline}
\frac{1}{v} \frac{\psi_{m,k+1/2}(x) - \psi_{m,k}(x)}{\Delta t/2} + \mu_m \frac{\partial \psi_{m,k+1/2}(x)}{\partial x} + \Sigma(x) \psi_{m,k+1/2}(x) \\
= \frac{1}{2} \left( \Sigma_{s}(x) \sum\limits_{n=1}^N w_n \psi_{n,k+1/2}(x) + Q_{ k+1/2}(x) \right) \;,
\end{multline}
\end{subequations}
where $\Delta t$ is the time step size, $k$ indicates time-average quantities, and $k\pm1/2$ indicates time-edge quantities.
Then, we discretize in space using simple corner balance, which involves a spatial integration over the right and left halves of a spatial cell:
\begin{subequations}
\label{eq:scb-mb}
\begin{multline}
\label{eq:scb-mb-a}
\frac{\Delta x_j}{2} \frac{1}{v} \left( \frac{\psi_{m,k+1/2,j,L} - \psi_{m,k-1/2,j,L}}{\Delta t} \right)
 + \mu_m \left[ \frac{\left( \psi_{m,k,j,L} + \psi_{m,k,j,R} \right)}{2}  - \psi_{m,k,j-1/2} \right] \\
+ \frac{\Delta x_j}{2} \Sigma_{j} \psi_{m,k,j,L} 
= \frac{\Delta x_j}{2} \frac{1}{2} \left( \Sigma_{s,j} \sum\limits_{n=1}^N w_n \psi_{n,k,j,L} + Q_{k,j,L} \right) \;,
\end{multline}  
\begin{multline}
\label{eq:scb-mb-b}
\frac{\Delta x_j}{2} \frac{1}{v} \left( \frac{\psi_{m,k+1/2,j,R} - \psi_{m,k-1/2,j,R}}{\Delta t} \right) +
\mu_m \left[ \psi_{m,k,j+1/2} - \frac{\left( \psi_{m,k,j,L} + \psi_{m,k,j,R} \right)}{2}   \right] \\
+ \frac{\Delta x_j}{2} \Sigma_{j} \psi_{m,k,j,R} = \frac{\Delta x_j}{2} \frac{1}{2} \left( \Sigma_{s,j} \sum\limits_{n=1}^N w_n \psi_{n,k,j,R} + Q_{k,j,R} \right) \;,
\end{multline}  
\begin{multline}
\label{eq:scb-mb-c}
\frac{\Delta x_j}{2} \frac{1}{v} \left( \frac{\psi_{m,k+1/2,j,L} - \psi_{m,k,j,L}}{\Delta t/2} \right) \\
+ \mu_m \left[ \frac{\left( \psi_{m,k+1/2,j,L} + \psi_{m,k+1/2,j,R} \right)}{2}  - \psi_{m,k+1/2,j-1/2} \right]
+ \frac{\Delta x_j}{2} \Sigma_{j} \psi_{m,k+1/2,j,L} \\
= \frac{\Delta x_j}{2} \frac{1}{2} \left( \Sigma_{s,j} \sum\limits_{n=1}^N w_n \psi_{n,k+1/2,j,L} + Q_{k+1/2,j,L} \right) \;,
\end{multline}    
\begin{multline}
\label{eq:scb-mb-d}
\frac{\Delta x_j}{2} \frac{1}{v} \left( \frac{\psi_{m,k+1/2,j,R} - \psi_{m,k,j,R}}{\Delta t/2} \right) + \\
\mu_m \left[ \psi_{m,k+1/2,j+1/2} - \frac{\left( \psi_{m,k+1/2,j,L} + \psi_{m,k+1/2,j,R} \right)}{2}   \right]
+ \frac{\Delta x_j}{2} \Sigma_{j} \psi_{m,k+1/2,j,R} \\
= \frac{\Delta x_j}{2} \frac{1}{2} \left( \Sigma_{s,j} \sum\limits_{n=1}^N w_n \psi_{n,k+1/2,j,R} + Q_{k+1/2,j,R} \right) \;,
\end{multline} 
\end{subequations}
where $\Delta x$ is the cell width, $j$ is the spatial cell index, $L/R$ is the left or right half cell, respectively.
These equations contain the first of the two simple spatial closures---the angular flux at the cell midpoint is a simple average of the two half-cell average quantities:
\begin{subequations}
\begin{equation}
  \psi_{m,k}(x_j) =  \frac{\left( \psi_{m,k,j,L} + \psi_{m,k,j,R} \right)}{2} \;,
\end{equation}
\begin{equation}
  \psi_{m,k+1/2}(x_j) =  \frac{\left( \psi_{m,k+1/2,j,L} + \psi_{m,k+1/2,j,R} \right)}{2} \;.
\end{equation}
\end{subequations}
The second closure is an \textit{upstream} prescription for the cell-edge angular flux:
\begin{subequations}
\begin{equation}
  \psi_{m,k,j+1/2} =
  \begin{cases}
  \psi_{m,k,j,R}, & \mu_m > 0, \\
  \psi_{m,k,j+1,L}, & \mu_m < 0 \;,
  \end{cases}
\end{equation}
\begin{equation}
  \psi_{m,k+1/2,j+1/2} =
  \begin{cases}
  \psi_{m,k+1/2,j,R}, & \mu_m > 0, \\
  \psi_{m,k+1/2,j+1,L}, & \mu_m < 0 \;.
  \end{cases}
\end{equation}
\end{subequations}
Figure~\ref{fig:stencil} shows the stencil location for angular flux and source terms. 

Solving Eqs.~\eqref{eq:scb-mb} iteratively requires operator splitting.
Unknown values (from the current iteration, noted by $(l+1)$) are moved to the left-hand side to form a large system of linear equations.
In SI, the scalar flux in the scattering source is evaluated at the previous iteration $(l)$, decoupling angles and coupling space.
OCI allows the fluxes incident to the cell---defined by upstream closures---to lag, thus decoupling cells from one another within an iteration.

In OCI, the scattering source is subtracted to the left-hand side and  prior iteration values are employed for all information incident on cell $j$ (moved to the right-hand side).
This means that all $4N$ angular fluxes ($N$ angles at $L$ and $R$, $k$ and $k+1/2$) are computed simultaneously in cell $j$.
This yields a linear system for each cell $j$:
\begin{equation}
    \label{eq:oci}
    \left( \bm{L}_{c,j} - \bm{S}_j \right) \Psi_j^{(l+1)} = -\textbf{L}_{b,j} \Psi_j^{(l)} + \textbf{Q} \;, 
\end{equation}
where $l$ is the iteration index.
The right-hand side can be combined into a known vector
\begin{equation}
     \left( \bm{L}_{c,j} - \bm{S}_j \right) \Psi_j^{(l+1)}  = \bm{b}_j \; ,
\end{equation}
where $\bm{L}_{c,j}$ and $\bm{S}_j$ are both of size $4N\times4N$ and likewise $\bm{b}_j$ is a vector of length $4N$.
\newcommand{\lcmj}[1]{\begin{bmatrix} \bm{L}_{c,j,#1} \end{bmatrix} }
\newcommand{\zeros}{\begin{bmatrix} 0 \end{bmatrix} }
The within-cell operator is
\begin{subequations}
\begin{equation}
    \label{eq:Aja}
    \bm{L}_{c,j} = \begin{bmatrix}
        \lcmj{1} &  &  &  &  \\
          & \ddots  &  &  & \\
          &  & \lcmj{m} &  & \\
          &  &  & \ddots &  \\
          &  &  &  & \lcmj{N}
    \end{bmatrix} \;,
\end{equation}
with zeros elsewhere, where
\begin{equation} \bm{L}_{c,j,m} =
    \label{eq:Aj}
    \begin{bmatrix}
    \frac{|\mu_m| + \Delta x_j \Sigma_{j} }{2}  & \frac{\mu_m}{2} & \frac{\Delta x_j}{2 v \Delta t} & 0 \\
    - \frac{\mu_m}{2} & \frac{|\mu_m| +  \Delta x_j \Sigma_{j,g} }{2} & 0 & \frac{\Delta x_j}{2 v \Delta t} \\
    -\frac{\Delta x_j}{v \Delta t}  & 0 & \frac{\Delta x_j}{v \Delta t} + \frac{|\mu_m| + \Delta x_j \Sigma_{j,g} }{2}  & \frac{\mu_m}{2}  \\
    0 &  -\frac{\Delta x_j}{v \Delta t}  &  - \frac{\mu_m}{2} & \frac{\Delta x_j}{v \Delta t}+ \frac{|\mu_m| + \Delta x_j \Sigma_{j,g}}{2}  \\
    \end{bmatrix} \; .
\end{equation}
\end{subequations}
The right-hand side is
\begin{subequations}
\label{eq:rhs_sg}
\begin{equation}
    \bm{b}_j =  \left[
    \bm{b}_{j,1} \; \bm{b}_{j,2} \; \cdots \; \bm{b}_{j,N} \right]^{T} \;.
\end{equation}
As the linear system in each cell contains contributions from all angles (both positive and negative) $b_{j,m}$ is given by
\begin{equation} 
    \bm{b}_{j,m} = 
    \begin{cases}
        \bm{b}_{j,m}^+ & \mu_m>0 \\
        \bm{b}_{j,m}^- & \mu_m<0 \\
    \end{cases} \;,
\end{equation}
where
\begin{equation}
     \bm{b}_{j,m}^+ = \begin{bmatrix}
     \frac{\Delta x_j}{4} Q_{k,j,L} + \frac{\Delta x_j}{2 v \Delta t} \psi_{m,k-1/2,j,L} + \mu_m \psi^{(l)}_{m,k,j-1,R} \\
     \frac{\Delta x_j}{4}Q_{k,j,R} + \frac{\Delta x_j}{2 v \Delta t} \psi_{m,k-1/2,j,R} \\
     \frac{\Delta x_j}{4}Q_{k+1/2,j,L} + \mu_m \psi^{(l)}_{m,k+1/2,j-1,R} \\
     \frac{\Delta x_j}{4} Q_{k+1/2,j,R} 
    \end{bmatrix} \;,
\end{equation}
and
\begin{equation}
    \bm{b}_{j,m}^- = \begin{bmatrix}
    \frac{\Delta x_j}{4}  Q_{k,j,L} + \frac{\Delta x_j}{2 v \Delta t} \psi_{m,k-1/2,j,L}  \\
    \frac{\Delta x_j}{4}  Q_{k,j,R} + \frac{\Delta x_j}{2 v \Delta t} \psi_{m,k-1/2,j,R} - \mu_m \psi^{(l)}_{m,k,j+1,L}  \\
    \frac{\Delta x_j}{4}  Q_{k+1/2,j,L}  \\
    \frac{\Delta x_j}{4}  Q_{k+1/2,j,R} - \mu_m \psi^{(l)}_{m,k+1/2,j+1,L}
    \end{bmatrix} \;.
\end{equation}
\end{subequations}
The elements of the ${S_j}$ matrix are defined by
\begin{equation}
    \label{eq:scatter}
   [\mathbf{S}_j]_{k,l} = \begin{cases}
			\frac{\Delta x_j \Sigma_{s,j}}{4}w_{\frac{|(r-s)|}{3}}, & \text{if $\mod{\frac{(r-s)}{3} =0}$}\\
            0, & \text{otherwise}
		 \end{cases} \; ,
\end{equation}
where $w$ are the angular quadrature weights, and $r$ and $s$ are the rows and columns of the scattering matrix, respectively.
Finally,
\begin{subequations}
\label{eq:x_vec}
\begin{equation}
\Psi^{(l+1)}_j =  \begin{bmatrix}
    \bm{\psi}_{j,1}^{(l+1)}, \;
    \bm{\psi}_{j,2}^{(l+1)}, \;
    \cdots \;
    \bm{\psi}_{j,N}^{(l+1)}
    \end{bmatrix} ^{T} \; ,
\end{equation}
where
\begin{equation} 
\bm{\psi}_{j,n}^{(l+1)} = \begin{bmatrix}
    \psi_{n,k,j,L}^{(l+1)}, \;
    \psi_{n,k,j,R}^{(l+1)}, \;
    \psi_{n,k+1/2,j,L}^{(l+1)}, \;
    \psi_{n,k+1/2,j,R}^{(l+1)}
    \end{bmatrix}^{T} \;.
\end{equation}
\end{subequations}
One-cell inversion iterations continue until
\begin{equation}
    ||\Psi^{(l+1)}-\Psi^{(l)}||_{2} < \epsilon(1-\rho_e) \; ,
\end{equation}
where $\epsilon$ is the convergence tolerance and
\begin{equation}
    \rho_e = \frac{||\Psi^{(l+1)}-\Psi^{(l)}||_{2}}{||\Psi^{(l)}-\Psi^{(l-1)}||_{2}} \; ,
\end{equation}
is an empirical estimation of the spectral radius computed at every iteration of a transport solve.
After convergence, the time-step counter increments and the time-step process can be repeated.

Generally, Jacobi and Gauss--Seidel iterations converge faster when a system is more diagonally dominant \cite{isaacson_numerical_1966, golub_matrix_1983}.
Equation~\eqref{eq:Aj} contains ($ \delta/2 = \Delta x\Sigma/2 $) on the diagonals.
So in the thin limit (when $\delta\rightarrow 0$) the system becomes overall less diagonally dominant and converges more slowly.
However Equation~\eqref{eq:Aj} also involves $\Delta x/(v\Delta t)$ terms in elements (3,3) and (4,4).
Thus, a smaller time step will cause the system to become more diagonally dominant.
We provide a similar description of simple corner balance and multiple balance time discretization for an unpreconditioned source iteration \cite{morgan2023oci}.

\subsection{Fourier analysis: time-stepping scheme}

To ensure that the combination of higher-order discretization schemes remains an unconditionally stable time-marching method, we perform Fourier analysis (also known as Von Neumann stability analysis)~\cite{leveque2007finite}.
A time marching scheme
\begin{equation}
    \Psi_{k+1/2} = \bm{K} \Psi_{k-1/2} \;,
\end{equation}
where $\bm{K}$ is the time iteration matrix, is unconditionally stable when the Von Neumann stability condition is met:
\begin{equation}
    \sup(|\lambda_{K}|) \leq 1 \;,
    \label{eq:unconstab}
\end{equation}
where $\lambda_{K}$ are all the eigenvalues of $\bm{K}$ \cite{golub_matrix_1983, isaacson_numerical_1966}.
$\bm{K}$ can be derived for a given model problem.
We consider a model problem consisting of a homogeneous infinite medium with no scattering to derive the eigenfunction of the time-dependent multiple balance, simple corner balance discretization scheme.
Since this problem has no scattering, each angle can be solved independently of every other angle, and no operator splitting is required.
We first start by describing the absolute error of the angular flux at time step $(k+1/2)$:
\begin{equation}
    \mathbf{f}_{k+1/2} = \Psi_{\text{exact}} - \Psi_{k+1/2} \;.
\end{equation}
We then assume that each unknown representing the error on our discretization stencil is expanded in a series of temporal Fourier modes, where each mode has a coefficient ($a$, $b$, $c$, or $d$), an amplitude ($\lambda$), and a shape function ($e^{i\omega x}$).
We can then define a Fourier ansatz for the error propagated through a time step:
\begin{subequations}
\begin{align}
    f_{k+1/2,j,L} &= \lambda^{k+1}a e^{i\omega j} \; ,
    &
    f_{k+1/2,j,R} &= \lambda^{k+1}b e^{i\omega j} \; ,
\end{align}
\begin{align}
    f_{k,j,L} &= \lambda^{k}c e^{i\omega j} \; ,
    &
    f_{k,j,R} &= \lambda^{k}d e^{i\omega j} \; ,
\end{align}
\begin{align}
    f_{k-1/2,j,L} &= \lambda^{k}a e^{i\omega j} \; ,
    &
    f_{k-1/2,j,R} &= \lambda^{k}b e^{i\omega j} \; ,
\end{align}
\end{subequations}
where $k$ is the time step, $i=\sqrt{-1}$, $\lambda$ is the eigenvalue, $\omega$ is the wave number, and $j$ is cell index. 
Substituting our ansatz into the error form of Eqs.~\eqref{eq:scb-mb} and assuming $\mu>0$, we simplify to form
\begin{subequations}
\begin{equation}
    \label{eq:eig1}
    \frac{\Delta x}{2} \frac{1}{v \Delta t} (\lambda a- a) + \mu \left(\frac{c+d}{2} - d e ^{-i\omega} \right) + \frac{\Sigma\Delta x }{2}c = 0 \; ,
\end{equation}
\begin{equation}
\label{eq:eig2}
    \frac{\Delta x}{2} \frac{1}{v \Delta t} (\lambda b - b) + \mu \left( ce^{i\omega} - \frac{c+d}{2} \right) + \frac{\Sigma\Delta x }{2}d = 0 \; ,
\end{equation}
\begin{equation}
\label{eq:eig3}
    \frac{\Delta x}{2} \frac{2}{v \Delta t} (\lambda a - c) + \lambda\mu \left( \frac{a+b}{2} - b e^{-i\omega} \right) + \frac{\Sigma\Delta x }{2}a \lambda = 0 \; ,
\end{equation}
\begin{equation}
\label{eq:eig4}
    \frac{\Delta x}{2} \frac{2}{v \Delta t} (\lambda b - d) + \lambda\mu \left( d e^{i\omega} + \frac{c+d}{2} \right) + \frac{\Sigma\Delta x}{2} b\lambda = 0 \;.
\end{equation}
\end{subequations}
Next, we combine Eq.~\eqref{eq:eig1} into \eqref{eq:eig2}:
\begin{equation}
    \label{eq:f_a-1}
    \begin{bmatrix}
        c\\d
    \end{bmatrix}
    = \bm{K_{+}}^{-1} \frac{\Delta x}{2}\frac{1}{v\Delta t}(1-\lambda)
    \begin{bmatrix}
        a\\b
    \end{bmatrix} \;,
\end{equation}
where
\begin{equation}
    {\bm{K_{+}}} = 
    \begin{bmatrix}
    \frac{\mu}{2}+\frac{\Sigma \Delta x}{2} & \mu (\frac{1}{2} - e^{-i\omega}) \\
    -\frac{\mu}{2} & \frac{\mu}{2}+\frac{\Sigma \Delta x}{2}
    \end{bmatrix} \;.
\end{equation}
Then, doing the same with Eq.~\eqref{eq:eig3} into \eqref{eq:eig4}:
\begin{equation}
    \label{eq:f_a-2}
    \lambda \left( \bm{K_{+}} + \frac{\Delta x}{v \Delta t} \bm{I} \right)  \begin{bmatrix}
        a \\ b
    \end{bmatrix} = \frac{\Delta x}{v\Delta t} \begin{bmatrix}
        c \\ d
\end{bmatrix} \; ,
\end{equation}
where $\bm{I}$ is the identity matrix. Combining Eq.~\eqref{eq:f_a-1} into \eqref{eq:f_a-2} gives
\begin{equation}
    \lambda
    \begin{bmatrix}
        a \\b
    \end{bmatrix}
    = \left[ \bm{K_{+}} + \frac{\Delta x}{v\Delta t} \bm{I} + \gamma\bm{K_{+}}^{-1} \right]^{-1} \gamma \bm{K_{+}}^{-1}
    \begin{bmatrix}
        a \\ b
    \end{bmatrix} \; ,
\end{equation}
where $\gamma = \frac{\Delta x}{v\Delta t}  \frac{\Delta x}{2v\Delta t}$.
This can then be more appropriately posed as an eigenfunction:
\begin{equation}
    \lambda_K
    \bm{a}
    = \bm{K} \bm{a}\;,
\end{equation}
where
\begin{equation}
    \bm{K} = \gamma \left( \bm{K_{+}}\bm{K_{+}} + \frac{\Delta x}{v \Delta t}\bm{K_{+}} + \gamma \bm{I}\right)^{-1}
\end{equation}
and the eigenvector is
\begin{equation}
    \bf{a} = \begin{bmatrix}
        a, & b
    \end{bmatrix} ^T \;.
\end{equation}
The analysis is similar for $\mu<0$ only with
\begin{equation}
    \bm{K_{-}} = \begin{bmatrix}
        -\frac{\mu}{2} + \frac{\Sigma\Delta x}{2} & \frac{\mu}{2} \\
        \mu \left ( e^{i\omega} - \frac{1}{2} \right) & -\frac{\mu}{2} + \frac{\Sigma\Delta x}{2} 
    \end{bmatrix} \;.
\end{equation}

This system can be numerically solved after making discrete selections of $\mu \in [-1, 1]$ and $\omega \in (0,2\pi]$ at a point in the perameter space ($\Delta x$, $\Delta t$, $v$, $\Sigma$) with \texttt{numpy.max(numpy.abs(numpy.linalg.\\eig(K)))}~\cite{harris2020array}.

\begin{figure}
    \centering
    \includegraphics[width=0.9\linewidth]{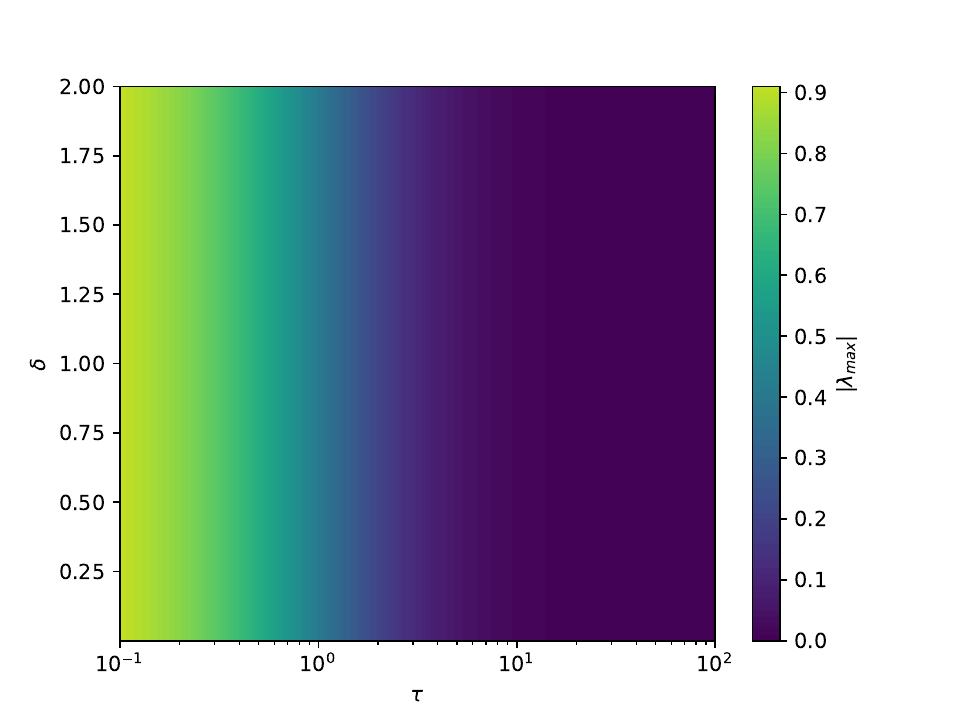}
    \caption{$|\lambda_{\max}|$ from numerically solved multiple balance time discretization and simple corner balance Fourier system over choices in mean free time ($\tau$) and cellular optical thickness ($\delta$) in S$_{16}$.}
    \label{fig:mb-scb}
\end{figure}

Figure \ref{fig:mb-scb} shows the absolute value of the maximum eigenvalues of $\bm{K}$ at various points in mean free time ($\tau=\Sigma v\Delta t$) and cellular optical thickness ($\delta=\Sigma\Delta x$) at 75 discrete points $\omega \in (0,2\pi]$ in S$_{16}$.
None of $|\lambda_{max}|$ are above one, which means the Von Neumann stability criterion in Eq. \eqref{eq:unconstab} is satisfied and the combination of the multiple balance time discretization and the simple corner balance scheme is unconditionally stable for this infinite homogeneous medium problem with no scattering.

\subsection{Fourier analysis: OCI iterative scheme}
\label{sec:methods-faoci}
\newcommand{\exi}{e^{i\lambda\Sigma x_j}}
\newcommand{\omlp}{\omega^{(l+1)}}
\newcommand{\oml}{\omega^{(l)}}
\newcommand{\dx}{\Delta x}
\newcommand{\dt}{\Delta t}
\newcommand{\scatsum}{\sum^{M}_{n=0}}

To study the impact of time dependence on the convergence of an OCI iteration, we conduct a Fourier analysis on the error equation of an infinite-homogeneous medium model problem in slab geometry in a single time step.
Similar to the analysis in the previous section, we can assert that for an iteration scheme
\begin{equation}
    \Psi^{(l+1)} = \bm{T} \Psi^{(l)} \;,
\end{equation}
where $(l)$ is the iteration counter, convergence rate is
\begin{equation}
   \rho = \sup(|\bm{\lambda}_{\bm{T}}|)\;,
\end{equation}
 where $\lambda_{\bm{T}}$ contains the eigenvalues of $\bm{T}$ \cite{golub_matrix_1983, isaacson_numerical_1966}.
An iterative method will converge if and only if $\rho<1$. 
Furthermore, iterations converge faster for smaller $\rho$.

To derive the transport matrix $\bm{T}$ we can again use Fourier separation analysis on a model problem.
We first start by describing the absolute error of the angular flux at iteration step $(l)$
\begin{equation}
    \mathbf{f}^l = \Psi^{\text{converged}} - \Psi^l \;,
\end{equation}
and our Fourier ansatz on a functional form of that error
\begin{subequations}
    \label{eq:anz}
    \begin{align}
        f^{(l)}_{m,k,j,L/R} &= \omega^{(l)}a_{m,L/R}e^{i\lambda\Sigma x_j} \; ,
        &
        f^{(l)}_{m,k+1/2,j,L/R} &= \omega^{(l)}b_{m,L/R}e^{i\lambda\Sigma x_j} \;.
    \end{align}
\end{subequations} 
The upstream closures at the left boundary of the cell are
\begin{subequations}
\begin{equation}
    f_{m,k,j-1/2} =
    \begin{cases}
        f_{m,k,j-1, R} \;, & \mu > 0 \\
        f_{m,k,j, L} \;, & \mu < 0 
    \end{cases} \;,
\end{equation}
and at the right are
\begin{equation}
    f_{m,k,j+1/2} =
    \begin{cases}
        f_{m,k,j, R} \;, & \mu > 0 \\
        f_{m,k,j+1, L} \;, & \mu < 0 
    \end{cases} \;.
\end{equation}
\end{subequations}

Now, substitute the ansatz and upstream closures into the error form of Eq. \eqref{eq:scb-mb} and derive the eigensystem.
This is done by (1) collecting like terms, (2) dividing both sides by $\omega^{(l)} e^{i\Sigma x_j}$, (3) isolating terms with a remaining $\omega$ to the left-hand side, and finally (4) forming the eigensystem into the iteration matrix over all angular directions:
\begin{equation}
    \bm{T}_{OCI} = \left( 
    \bm{L}_c
    - \bm{S}
    \right)^{-1}
    \begin{bmatrix}
        \bm{L}_b^- & 0\\
        0 & \bm{L}_b^+
    \end{bmatrix} \;,
\end{equation}
which now forms a well-posed eigenvalue problem over all angles:
\begin{equation}
    \lambda\bm{a} = \bm{T} \bm{a} \; ,
\end{equation}
where the eigenvector $\mathbf{a}$ is defined by
\begin{equation}
    \mathbf{a} = \begin{bmatrix}
        a_{1} & a_{2} & \cdots & a_M
    \end{bmatrix} ^T \;,
\end{equation}
\begin{equation}
    a_m = \begin{bmatrix}
        a_{mR} & a_{mL} & b_{mR} & b_{mL} 
    \end{bmatrix} ^T \; ,
\end{equation}
$\bm{L}_c$ is the linear within-cell transport operator defined by Eqs.~\eqref{eq:Aja} and \eqref{eq:Aj}, 
\begin{equation}
    \bm{L}_b^+ = 
    \begin{bmatrix}
        0 & 0 & 0 & 0 \\
        -\mu_m e^{-i\lambda\sigma\dx} & 0 & 0 & 0 \\
        0 & 0 & 0 & 0 \\
        0 & 0 & -\mu_m e^{-i\lambda\sigma\dx} & 0
    \end{bmatrix} \; ,
\end{equation}
and
\begin{equation}
    \bm{L}_b^- = 
    \begin{bmatrix}
        0 & \mu_m e^{i\lambda\sigma\dx} & 0 & 0 \\
        0 & 0 & 0 & 0 \\
        0 & 0 & 0 & \mu_m e^{i\lambda\sigma\dx} \\
        0 & 0 & 0 & 0
    \end{bmatrix} \; .
\end{equation}
The scattering matrix is again akin to the previously described transport matrix in Eq.~\eqref{eq:scatter}.
Finally, to numerically evaluate the spectral radius we form the system for a given set of angles and weights from Gauss--Legendre quadrature and solve with \texttt{numpy.max(numpy.abs(numpy.linalg.eig(T)))} for $\omega \in [0,2\pi]$ at discrete points.
We vary the cellular optical thickness ($\delta =\Sigma \Delta x$), mean free time ($\tau = \Sigma v\Delta t$), and scattering ratio ($c=\Sigma_s/\Sigma$) to study convergence behavior in various physical regimes. 
The analogous eigensystem for source iteration is
\begin{equation}
    \bm{T}_{SI} = \left( 
    \bm{L}_c
    + \begin{bmatrix}
        \bm{L}_b^- & 0\\
        0 & \bm{L}_b^+
    \end{bmatrix}
    \right)^{-1}
    \bm{S} \; .
\end{equation} 
Section~\ref{sec:results-faoci} contains the results of this analysis.

\subsection{OCI multi-group transport}

We extend our single-energy derivations presented in Section~\ref{sec:methods-derv} to be energy dependent. 
Elements of the $\mathbf{S_{g' \rightarrow g}}$ matrix are now defined by
\begin{equation}
    \label{eq:scatter_mg}
   [\mathbf{S}_{g' \rightarrow g,j}]_{k.l} = \begin{cases}
			\frac{\Delta x_j \Sigma_{s,g'\rightarrow g,j}}{4}w_{\frac{|(r-s)|}{3}}, & \text{if $\mod{\frac{(r-s)}{3} =0}$}\\
            0, & \text{otherwise}
		 \end{cases} \; ,
\end{equation}
where $g' \rightarrow g$ indicates transfer from group $g'$ to group $g$ and $w$ are the quadrature weights. 
\begin{subequations}
The full system of linear equations in all groups and angles in cell $j$ becomes
\begin{equation}
    \bm{A}_j \bm{\Psi}_j = \bm{b}_j \; ,
\end{equation}
where
\begin{equation}
    \label{eq:fullOCIlhs}
    \bm{A}_j = 
    \begin{bmatrix}
        \bm{L}_{c,j,1} -\bm{S}_{1\rightarrow1,j} & -\bm{S}_{2\rightarrow1,j} & \cdots & \cdots& -\bm{S}_{G\rightarrow1,j}\\
        -\bm{S}_{1\rightarrow2,j} & \ddots & & & \vdots\\
         \vdots & & \bm{L}_{c,j,g}-\bm{S}_{g\rightarrow g,j} &  & \vdots\\
        \vdots & &  &  \ddots & \vdots \\
        -\bm{S}_{1\rightarrow G,j} & \cdots & \cdots & & \bm{L}_{c,j,G} -\bm{S}_{G\rightarrow G,j}
    \end{bmatrix} \; ,
\end{equation}
and Eqs.~\eqref{eq:rhs_sg} and \eqref{eq:x_vec} are extended to multi-group by
\begin{equation}
    \label{eq:fullOCIrhs}
    \bm{b}_j = 
    \begin{bmatrix}
        b_{j,1}, & b_{j,2}, & \cdots, & b_{j,G}
    \end{bmatrix} ^T 
\end{equation}
and
\begin{equation}
    \bm{\Psi}_j = 
    \begin{bmatrix}
        \Psi_{j,1}, & \cdots & \Psi_{j,g}, & \cdots, & \Psi_{j,G}
    \end{bmatrix} ^T \; ,
\end{equation}
 with otherwise similar structure.
\end{subequations}

\begin{figure}
    \centering
    \includegraphics[width=.9\textwidth]{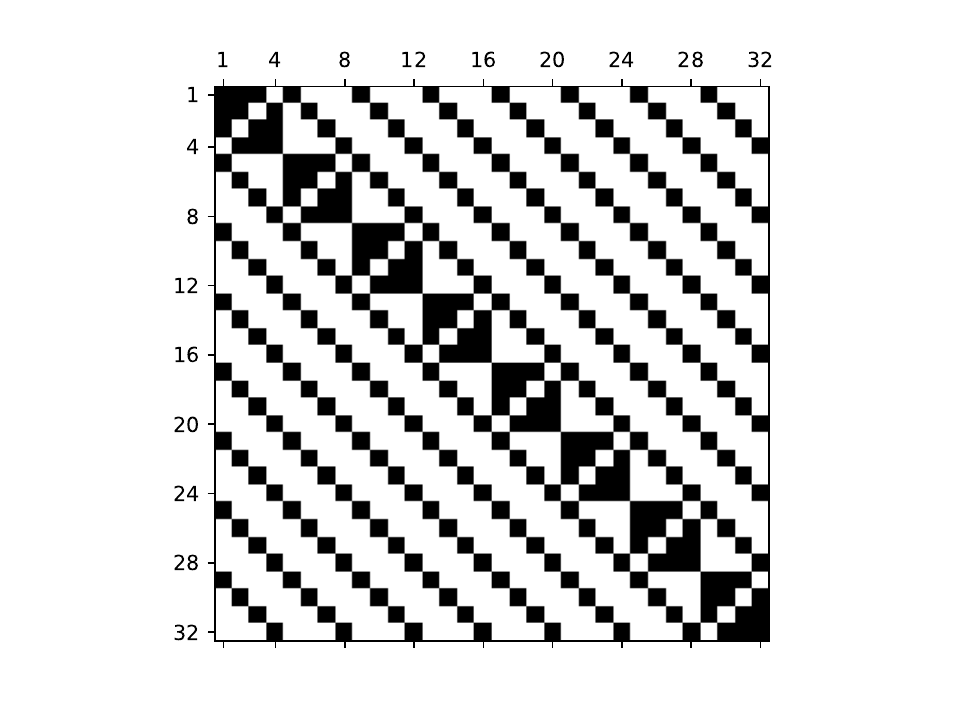}
    \caption{Sparsity pattern of a two group, four angle OCI $\bm{A}_j$ system generated for each cell.}
    \label{fig:spyA}
\end{figure}

Figure~\ref{fig:spyA} shows the structure of the within-cell system of equations arises from a two-group four-angle problem.
While $A_j$ does have significant sparsity, with an occupancy ratio of
\begin{equation}
    O_c = \frac{G(N+2)}{4NG} 
    \; ,
\end{equation}
which approaches 25\% with large angular and group counts, in this work we use dense representations in each cell because the matrix memory size for 1D transport is not limiting.

\subsection{Implementation on GPUs}

Implementing the OCI and SI approaches on GPUs requires a numerical linear algebra solver library like LAPACK~\cite{laug}.
Many high-performance open-source linear algebra tools exist (e.g., Trillinos \cite{trilinos-website}, PETSc \cite{petsc-user-ref}, MAGMA \cite{magma}), but we chose a vendor-supplied package depending on the hardware target of choice.
Our target hardware is an AMD MI250X so we use the AMD ROCm compute library to solve the system of equations.
Modern GPU vendor-supplied LAPACK libraries often include a \texttt{batched} class of solvers,
which operate on a group of like-sized systems in unison and are optimized by the hardware vendors.
For example, LU decomposition with pivoting (a generic direct solver for a system of linear equations) used in this work, comes from RocSolver's \texttt{strided\_batched\_dgesv} \cite{rocsolver}.

We use direct solvers here because all systems are relatively small, with orders ranging between 4 and 100.
This makes the use of a batched implementation of LU decomposition with pivoting ideal.
Furthermore, LAPACK-type implementations of \texttt{\_gesv} (\textbf{GE}neral linear \textbf{S}ol\textbf{V}e)  automatically return the $L+U+D$ decomposition of a generic matrix $A$.
So, in subsequent iterations, this system can be back solved quickly (using LAPACK \texttt{\_getrs}).
In this mode for both SI and OCI, the only user-defined device kernels are the RHS vector builders which are already memory safe operations.

This software engineering design will increase the memory footprint of OCI and SI as the $LHS$ matrices are stored in memory.
This is acceptable for 1D transport but more optimization may be required when moving to 2D and 3D solvers.

Algorithm \ref{alg:si} describes the convergence loop for source iteration.
$\mathbf{L}_{c,j,g,m}$ and $\bm{d}_{j,g,m}$ matrices are always of dimension $4\times 4$ and 4, respectively, for our space time-discretization scheme and are defined in Appendix~\ref{app:source_iteration}.
The number of systems to solve changes with the number of angles ($N$), groups ($G$), and cells ($J$).
In this algorithm the number of systems solved in parallel at one spatial index $j$ is $N \times G$.
SI requires host-side dispatching in every cell to execute the sequential nature of the sweep.
We implemented this algorithm to do all available computing at once (negative sweeps are happening in unison with positive ones in all angles and groups).
The angle parallelism exploited for this implementation of SI on the GPU is enabled by the \texttt{strided\_batched} call itself.
Thus, the traditional loop over all angles is implemented by the solver, not explicitly by the user.
Group-to-group communication is done at the end of every iteration.
The first iteration calls the full \texttt{\_gesv} algorithm, which returns the solution of the system and the $L+U+D$ decomposition in $A$.
Subsequent iterations just perform a back substitution (\texttt{\_getrs}).
Profiling shows that host functions (including host$\rightarrow$device and device$\rightarrow$host communication) account for up to around $9\%$ of the runtime in the largest problems we considered.

\SetKwComment{Comment}{//}{ }
\DontPrintSemicolon
\begin{algorithm}
    \label{alg:si}
    build $\mathbf{L}_{c,j,g,m}$ for each cell, angle, and group \Comment*[r]{Eq. \ref{eq:app:si_sys}}
    
    move $\mathbf{L}_{c,j,g,m}$ to device

    $\beta$ = $4NG$ \Comment*[r]{offset to a cell} 

    $l = 0$ \Comment*[r]{iteration counter}

    converged = false

    \While{!converged}{

    build constant part of $\bm{d}_{j,g,m}$ in all cells, angles, and groups \Comment*[r]{Eq. \eqref{eq:app:si_rhs}}

    move constant part of $\bm{d}_{j,g,m}$ to device
        
        \For{j = 0 to $J$ \Comment*[r]{parallel over groups and angles}}{ 

            build variable part of $\bm{d}_{j,g,m}$ at cells $j$ and $J-j$  \Comment*[r]{on GPU Eq. \eqref{eq:app:si_rhs}}

            \If{l=0}{
                \Comment*[r]{LHS in, L+U+D out}
                $\Psi_j$ = \texttt{GPU\_strided\_batched\_dgesv}($\mathbf{L}_{c,j}$[$\beta^{2}j$],$\bm{d}_{j}$[$\beta j$])
            }\Else{
                \Comment*[r]{back substitution}
                $\Psi_j$ = \texttt{GPU\_strided\_batched\_dgetrs}($\mathbf{L}_{c,j}$[$\beta^2j$],$\bm{d}_{j}$[$\beta j$]) 
            }
            }

        move $\Psi$ to Host

        $\bm{\phi}^l =\sum_{n=1}^{N} w _n\Psi^{l}_{n}$ \Comment*[r]{Eq. \eqref{eq:app:si_sf}}

        $e=||\bm{\phi}^l - \bm{\phi}^{l-1}||_2$

        $\rho_e = e^l / e^{l-1}$

        \If{$e < \epsilon(1-\rho_e)$}{
            converged = true
        }

        $e^{l-1} = e^l$ 

        $l++$
        
        move $\Phi^l$ to Host

        communicate group to group \Comment*[r]{Eq. \eqref{eq:app:si_update}}

    }
    \vspace{1.5em}
    \caption{Source iteration algorithm implemented on GPU where $\bm\phi$ is scalar flux. Equations in Appendix~\ref{app:source_iteration}. Simplified for brevity.}
\end{algorithm}

Algorithm \ref{alg:ocigpu} describes OCI's on-GPU convergence loop.
We found OCI to be more sensitive to within-iteration optimizations.
In some cases (specifically in the thin limit) OCI may require significantly more iterations to converge.
For that reason, it is imperative that the OCI iteration take place entirely on the GPU.
Luckily, OCI's algorithm is simpler to implement on GPUs because group-to-group communication happens within the solved systems.
We implemented the following algorithm to do that: everything under the \texttt{while} loop is wholly contained on the GPU, requiring minimal device-to-host communication.

\begin{algorithm}
    
    build $\bm{A}_j$ in all cells and move to device \Comment*[r]{Eq.~\eqref{eq:fullOCIlhs}}

    build constant part of $\bm{b}_j$ in all cells and move to device \Comment*[r]{Eq.~\eqref{eq:fullOCIrhs}}

    $l = 0$ \Comment*[r]{iteration counter}

    converged = false

    \While{!converged}{
        \Comment*[r]{incident angular fluxes from previous iteration}
        \Comment*[r]{user-defined GPU kernel}
        build variable part of $\bm{b}_j$ in all cells \Comment*[r]{Eq.~\eqref{eq:fullOCIrhs}}

        \If{l=0}{
            \Comment*[r]{A in-out becomes the L+U+D decomp}
            $\Psi$ = \texttt{GPU\_strided\_batched\_dgesv}($A$,$b$)
        }\Else{
            \Comment*[r]{back substitution}
            $\Psi$ = \texttt{GPU\_strided\_batched\_dgetrs}($A$,$b$) 
        }

        $e=||\Psi^l - \Psi^{l-1}||_2$ \Comment*[r]{Done on GPU using rocBLAS dr2n}

        $\rho_e = e^l / e^{l-1}$ \Comment*[r]{spectral radius estimation}

        \If{$e < \epsilon(1-\rho_e)$\Comment*[r]{controlling for false convergence}} { 
            converged = true
        }

        $e^{l-1} = e^l$

        $b^{l-1} = b^l$

        $l++$
            
    }
    
    move $\Psi$ to host

    \vspace{1.5em}
    
    \caption{One-cell inversion algorithm implemented on GPUs. Simplified for brevity.}
    \label{alg:ocigpu}
\end{algorithm}

OCI's systems are represented as dense within a cell and built in a strided-batched configuration to take advantage of the block sparsity.
However, now systems within an iteration can be dispatched in unison.
Just as with the SI algorithm, the GPU strided batched solver implements the parallel loop over all cells.
The intra-iteration $b$-vector production kernels are the only user-defined device functions required in this algorithm. These are relatively simple to implement as they are thread-safe operations.

\section{Results}
\label{sec:results}

In this section, we show results that support our initial conjecture that OCI convergence accelerates, more than SI, in transient transport calculations with decreased time step sizes.
We also further analyze OCI's performance on AMD MI250X GPUs using batched LAPACK solvers on a highly scattering problem from literature at multiple time step sizes and cell width values.

\subsection{Fourier analysis: transient iterative convergence rate}
\label{sec:results-faoci}

To study the impact of transient conditions on OCI we solve the Fourier system for steady-state and time-dependent transport derived in Section~\ref{sec:methods-faoci} both for simple corner balance in space.
For all Fourier analyses we sample $\lambda \in [0,2\pi]$ at \num{250} points and use \texttt{numpy.max(numpy.abs(numpy.eig(}$T$\texttt{)))} to compute spectral radius at a given point in parameter space ($\delta$ ($\Sigma\Delta x$), $\tau$ ($\Sigma v\Delta t$), and $c$ ($\Sigma_s$/$\Sigma$)) in S$_{8}$ using Gauss--Legendre quadrature.

Table~\ref{table:difflimit} shows spectral radii produced from steady-state and transient OCI systems with various choices of mean free time ($\tau$), at various cellular optical thicknesses ($\delta$).
Steady-state predictions show the expected and previously published results that $\rho=1$ when $c=1$ regardless of $\delta$.
However, for the time-dependent system, $\rho<1$ regardless of the considered $\tau$ and $\delta$.
Furthermore, as $\tau$ shrinks and $\delta$ grows, $\rho$ dramatically decreases, approaching zero at the smallest $\tau$ and largest $\delta$.

\begin{table}
  \centering
  \begin{tabular}{@{}l c c c @{}} \toprule
    $\tau$ & $\delta=10.$ & $\delta=1.0$ & $\delta=0.1$ \\ \midrule
    SS  & \num{1.0000} & \num{1.0000} & \num{1.0000} \\
    10 & \num{0.99522} & \num{0.99952} & \num{0.99995} \\
    1  & \num{0.95323} & \num{0.99522} & \num{0.99952} \\
    0.1   & \num{0.64031} & \num{0.95321} & \num{0.99522} \\
    0.01 & \num{0.11177} & \num{0.63343} & \num{0.95351} \\
    \bottomrule
  \end{tabular}
  \caption{OCI spectral radius $\rho$ in the diffusive limit ($c=\Sigma_s/\Sigma =1.0$) from Fourier analysis at various mean free time ($\tau$) and cellular optical thickness ($\delta$) values. SS indicates steady state.} 
  \label{table:difflimit} 
\end{table}

\begin{figure}
    \centering
    \includegraphics[width=\textwidth]{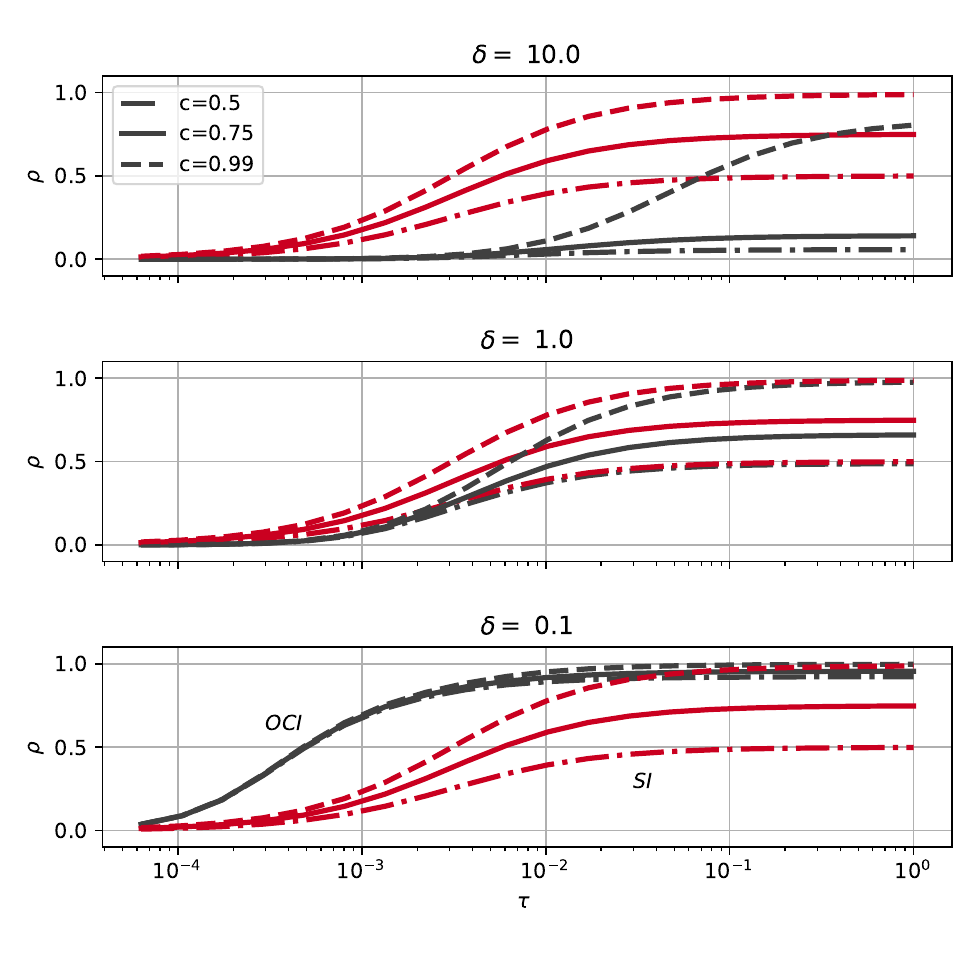}
    \caption{Spectral radius of OCI (black) and SI (red) over choices of mean free path ($\delta$), time step ($\Delta t$), and scattering ratio ($c$).}
    \label{fig:specrad_fa}
\end{figure}

Figure~\ref{fig:specrad_fa} shows $\rho$ predictions for OCI and SI produced from the Fourier system.
As previously published: as $\delta$ gets smaller, $\rho$ approaches 1 regardless of the scattering ratio.
As postulated in this work: as $\Delta t$ gets smaller, $\rho$ tends to 0---due to improvements in scattering ratio (which also affects SI) and increasing $\delta$---increasing the diagonal dominance of the iteration matrix.

Fourier analysis results also show that, depending on the location in parameter space, the dominant eigenvalue ($|\lambda_{max}|$) can have large imaginary components, with positive or negative real components and complex conjugate reflections over the real axis.
Complex dominant eigenvalues leading to oscillatory convergence patterns have previously been identified in spatial domain decomposition algorithms where $\rho=1$ when $\delta \rightarrow 0$ \cite{compeig2019ani}. 

\begin{figure}
    \centering
    \includegraphics[width=.49\textwidth]{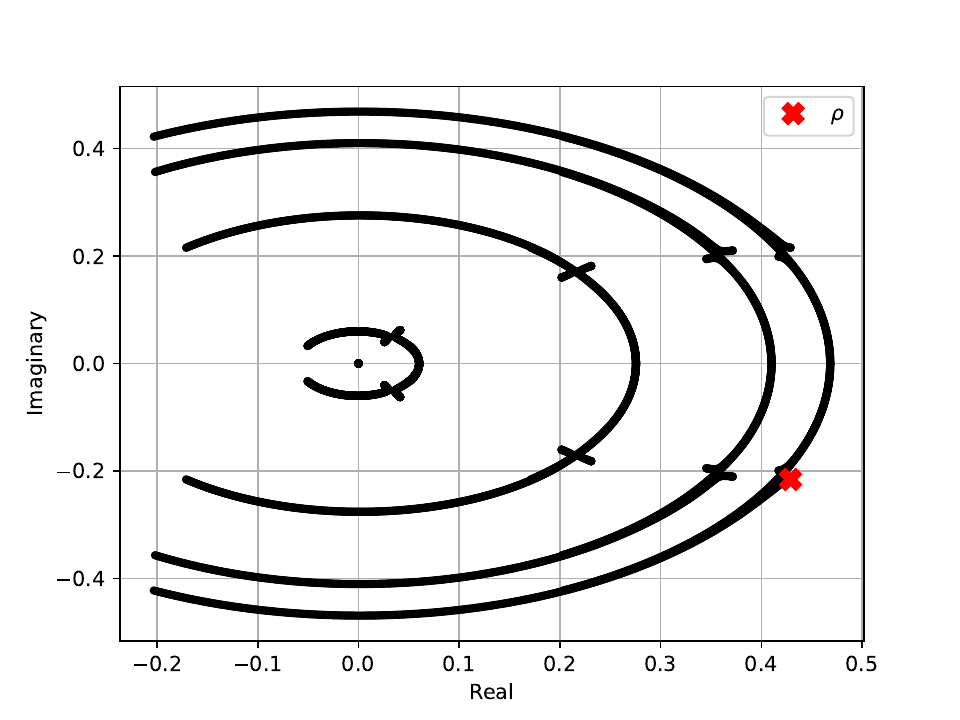}
    \includegraphics[width=.49\textwidth]{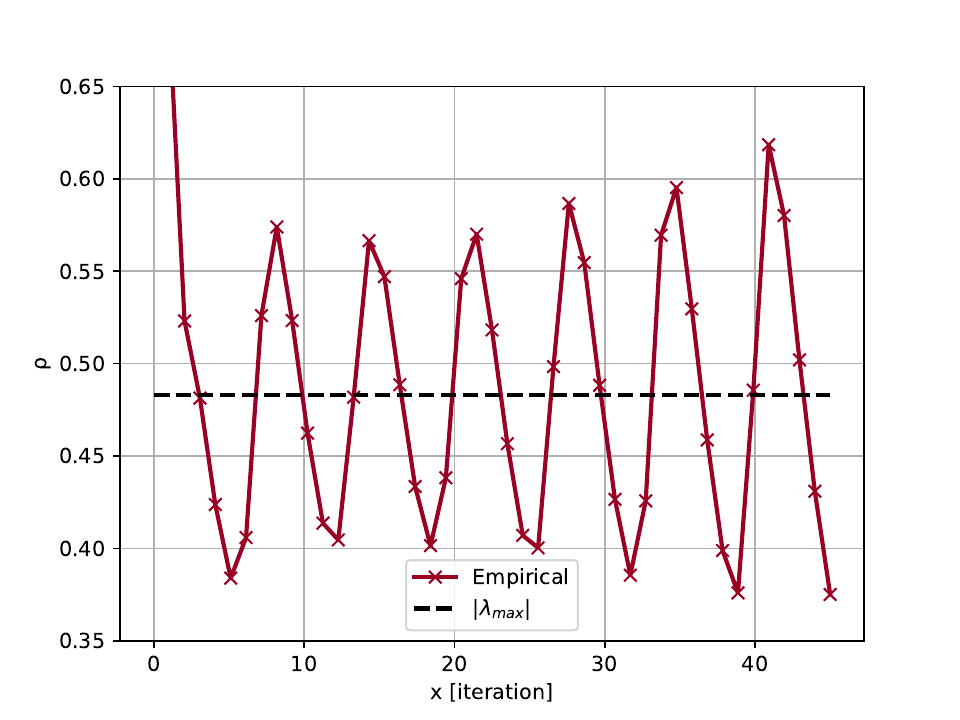}
    \caption{Eigenvalues in the complex plane of OCI (left), and spectral radius as a function of iteration from the empirical ratio of subsequent residuals and as predicted by Fourier analysis (right).}
    \label{fig:eigplot}
\end{figure}

Deterministic solvers are commonly verified against predictions of $\rho$ from Fourier analysis.
We attempted to do the same by running a problem with length \SI{100}{\centi\meter}, vacuum boundary conditions, a convergence tolerance of \num{1e-13}, $\Sigma=$ \SI{2.5}{\per\centi\meter}, $\Delta x=$ \SI{0.10}{\centi\meter}, $c=$ \num{0.9}, $\Delta t=$\SI{0.10}{\s}, $v=$ \SI{4.0}{\meter\per\s} ($\delta =$ \num{0.25}, $\tau=$ \num{1.0}), a random (uniform [0,1]) initial guess for the angular flux, and no material source in S$_8$.
The random initial guess excites all error modes and provides an anaclitic solution ($\Psi^{\text{converged}} = \bm{0}$) to compute iteration errors.
Figure \ref{fig:eigplot} on the left shows the predicted eigenvalues from Fourier analysis and indicates the dominant eigenvalue that contributes to $\rho$ for this particular problem.
In this case, that dominant eigenvalue has considerable real and complex components at $\lambda_{max} =$ \num{0.429} $+$ \num{0.216}$i$ and $\rho=$ \num{0.4831}.

Figure~\ref{fig:eigplot} on the right shows $\rho$ predicted from Fourier analysis (flat constant line) as well as $\rho$ measured from the ratio of subsequent residuals as a function of iteration count ($l$).
The empirically estimated value of $\rho$ oscillates around the predicted spectral radius until convergence, with a measured amplitude around \num{0.1}.
The oscillation of the empirically measured spectral radius also seems to grow through iteration count, which may be due to the compounding impact of truncation error and/or machine precision.
So, we cannot rigorously verify our implementation of OCI via Fourier results, because only a mean of the oscillation will match the only-real $\rho$ provided from Fourier results ($|\lambda_{max}|$).
More work is warranted to develop methods that can better capture the empirical behavior of the ratio of subsequent residuals produced from a transport solver and relate them to the complex dominant eigenvalues that may be predicted from Fourier analysis.

\subsection{Performance on GPUs}

Runtime results were gathered on the Tioga machine at Lawrence Livermore National Laboratory.
Tioga is an early access machine for LLNL's exascale-class El Capitan machine.
On its standard partition, Tioga's nodes have four AMD MI250X GPUs and one AMD EPYC 7A53 CPU.
Our methods are currently implemented for a single GPU, so this analysis will be limited to using a single graphics compute die of an MI250X.
We compiled using ROCm version 6.2.1 (includes rocSOLVER and rocBLAS libraries) and used double precision for all values represented.

To analyze performance, we adapt a test problem described by \cite{rosa_cellwise_2013} for a 1D time-dependent, multi-group problem.
Table~\ref{table:rosa_test} describes the material data for this 
two-group problem ($L=$ \SI{100}{\centi\meter}) with vacuum boundary conditions on either side. 
The initial condition is $\psi_{t=0} = 0$, and we analyze runtime performance over various choices of $\delta$ and quadrature order at time step sizes of $\Delta t=$ \SI{0.1}{\s} and \SI{10.0}{\s}.
The problem is highly scattering with a maximum scattering ratio of \num{0.99997}.

\begin{table}
  \centering
  \begin{tabular}{@{}c c c c c c@{}} \toprule
    Property & Group 1 & Group 2 & units \\ \midrule
    $\Sigma$ & 1.5454 &  0.45468 & cm$^{-1}$  \\
    $\Sigma_{s,g\rightarrow g}$  & 0.61789 &  0.0072534 & cm$^{-1}$  \\
    $\Sigma_{s,g'\rightarrow g}$  & 0.38211 &  0.92747 & cm$^{-1}$ \\
    $\Sigma_s/\Sigma$ & 0.99997 & 0.86012 & - \\
    $Q$ & 1 & 1 & cm$^{-3}$s$^{-1}$\\
    $v$ & 1 & 0.5 & cm s$^{-1}$ \\
    \bottomrule
  \end{tabular}
  \caption{Test problem material data and simulation parameters.}
  \label{table:rosa_test} 
\end{table}

\begin{figure}
    \centering
    \includegraphics[width=1.0\textwidth]{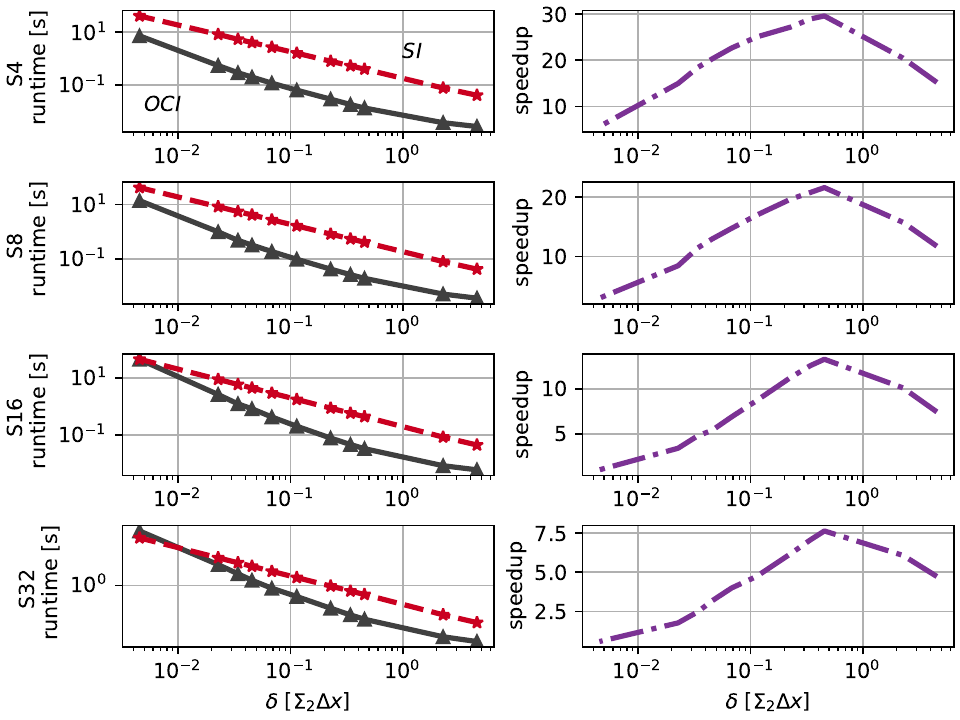}
    \caption{Wall-clock runtimes of the convergence loop (left) and 
    speedup of OCI over SI (right) at $\Delta t=$\SI{10.0}{\s} ($\tau=$ \num{2.2734}) as a function of $\delta$ and at various quadrature orders.}
    \label{fig:runtimes10.0}
\end{figure}

\begin{figure}
    \centering
    \includegraphics[width=1.0\textwidth]{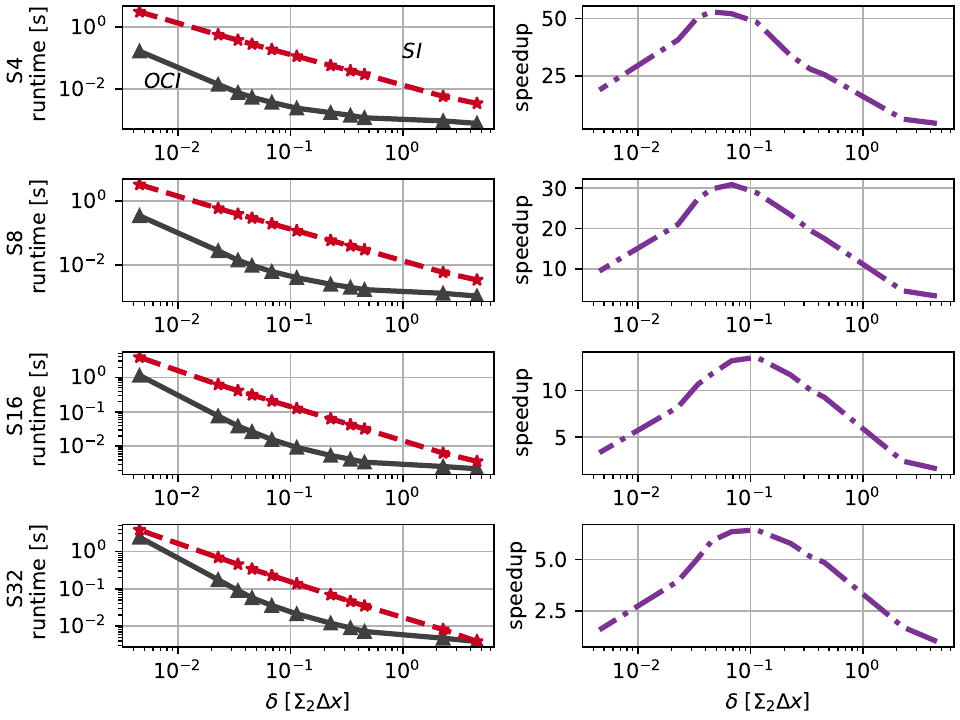}
    \caption{Wall-clock runtimes of the convergence loop (left) and 
    speedup of OCI over SI (right) at $\Delta t=$\SI{0.1}{\s} ($\tau=$ \num{0.0227}) as a function of $\delta$ and at various quadrature orders.}
    \label{fig:runtimes0.1}
\end{figure}

Figure~\ref{fig:runtimes10.0} on the left compares the wall clock runtime of OCI (in black) and SI (in red) over various selections of $\delta$ (controlled via $\Delta x$) with $\Delta t=$ \SI{10.0}{\s}, Figure~\ref{fig:runtimes10.0} on the right shows the speedup of OCI over SI.
In each row, we are increasing quadrature order to increase the overall dimensionality of the system.
Figure \ref{fig:runtimes0.1} shows the same information, but for $\Delta t=$ \SI{0.1}{\s}.
Runtimes are measured over the convergence loops (see Algorithms \ref{alg:si} and \ref{alg:ocigpu}), so do not include the building and moving the $A_j$ matrices from host to device.
The total cross section used in the $\delta$ scale is the limiting value (the smallest) from group 2 (see Table \ref{table:rosa_test}).

SI's convergence loop runtime increases linearly as cellular optical thickness decreases as there are more cells to solve  in serial.
The number of iterations required to converge the solution is the same but the size of the solution grows.
SI only has $NG$ $4\times4$ systems to solve at any moment so the amount of serial work increases with the number of cells (decreasing $\delta$).
However, as we increase quadrature order the runtime performance of SI actually improves because the solver has more parallelizable degrees of freedom.

For larger time steps, OCI shows less speed-up over SI as it slows for S$_{16}$ and S$_{32}$ quadratures in the thin limit.
The parallelizable degrees of freedom increase with the number of cells (by decreasing $\delta$), but the spectral radius decreases dramatically as cells get thinner.
In the thin limit, OCI requires more iterations to converge the solution, but those iterations can be done faster on the GPU than with SI.
OCI seems to have a ``sweet spot'', where the size of the matrices is optimal for the solver, before the spectral radius degrades in the thin limit.
This is observed at around $\delta=4$ for $\Delta t=$ \SI{10}{\s} and $\delta=0.1$ for $\Delta t=$ \SI{0.1}{\s}.
The location of this optimality depends on factors including optimizations at the solver level employed when compiling the vendor-supplied LAPACK libraries \cite{rocsolver}.
The smaller time step increases OCI's relative performance over SI, generally increasing speedup by upwards of 40\% for this highly scattering problem.

\section{Discussion}

The 1D convergence trends we present here agree with
previously published 2D steady-state Fourier results for OCI schemes 
(i.e., $\rho\rightarrow1$ as $\delta\rightarrow0$) \cite{rosa_cellwise_2013, man1994parallel}.
This leads us to expect that the relationship between mean free time and spectral radius will persist in higher spatial dimensions, 
but exactly how much dynamic impacts to OCI decrease $\rho$ in 2D transport has yet to be shown.

Parallel sweeping algorithms may not be well suited to GPUs where non-uniform work distributions come with significant overhead. 
Available results for full parallel sweeps on GPUs show that even optimized applications underperform relative to the theoretical hardware resources available \cite{Thomas_2024_profiling, wolfe2022roofline, kunen_kripke_2015, womeldorff_taking_2017, zerr_partisn_2019}.
On the other hand, space-parallel OCI uses the same parallel scheme in 2D and 3D as it does in 1D, with arithmetically intense operations that align well with the GPU parallelism paradigm.
So, we hypothesize that OCI can better take advantage of the compute resources available on GPUs in higher dimensions than full-parallel-sweep SI on GPUs, but this requires further study.

Production SI codes that simulate high-fidelity 3D models never form the actual iteration matrix, resulting in significant memory efficiency~\cite{partisn, evans_denovo_2010, kunen_porting_2019, MINARET, wang_necp_hydra_2020, ShemonEmilyR.2016PUM}.
Our implementation of OCI is not memory efficient, because we store the strided batched representation of the entire linear system for the iteration algorithm.
While this block-sparse representation does dramatically save memory, more optimizations to OCI may be needed when moving to 3D.
For example, OCI forms the iteration matrix as a linear operator.
Many algorithms exist to solve systems of linear operators without ever forming the matrix.
Libraries already exist to do this on CPUs and GPUs (e.g., LibCEED \cite{libceed-joss-paper}).
Even more simple optimization may involve using sparse representations for the cell blocks themselves and sparse direct solvers.

OCI algorithms may be better suited for domain-decomposed simulations (required for distributed-memory parallelism) than SI.
Some KBA algorithms can degrade with large processor counts \cite{baker_sn_2017}.
For a hypothetical OCI algorithm, communication between subdomains can be very organized, occurring at every iteration (or after a specified number of iterations) at the same time as all other subdomains.
This may lead to superior weak scaling across large decomposed problems on distributed (MPI) type systems.
Many parallel domain decomposition algorithms are already based on the same underlying parallel block Jacobi, Gauss--Seidel, or red-black iterations as OCI \cite{anistratov_iterative_2015, compeig2019ani}.

While this work is limited to AMD GPUs and the ROCm software stack, our implementation of OCI is portable to other GPU types.
In fact, the workflow present in OCI---solving many small dense or sparse linear algebra systems in parallel---is common in much numerical software.
We expect that these types of solvers will exist when porting to a new hardware accelerator used for scientific computing.
For example, Nvidia's cuSOLVER \cite{cuSovler} and Intel's oneMKL \cite{oneMKL} both implement strided batched versions of LU decomposition with pivoting for their GPUs.
Porting the OCI algorithm presented in Algorithm~\ref{alg:ocigpu} to another GPU would just require altering names of solver function calls, compiler commands, and switching out library header imports.

OCI algorithms may also be well suited for modeling anisotropic scattering distributions because all angles are computed at once in every cell.
On unstructured meshes, OCI algorithms avoid one challenge for sweep-based methods: when groups of cells have cyclic dependencies (i.e., when an incident transport angle is parallel to a cell boundary).

Regardless of how well an implementation of any OCI scheme performs, the inability to converge problems in the thin limit regardless of scattering ratio will continue to lead to lackluster performance in some problems.
In this work, we compared unpreconditioned SI to unpreconditioned OCI using fixed-point iterations.
When in production, SI typically uses a well-accepted set of acceleration schemes/preconditioners (most popularly diffusion synthetic acceleration) accompanied by Krylov subspace methods.
Likewise, some acceleration/preconditioning or Krylov methods may exist that can help OCI more-rapidly converge in the thin limit, while not significantly degrading the space-parallel performance of OCI.

Acceleration schemes for OCI have previously been explored, including transport synthetic acceleration \cite{tsa2009rosa} and using hybrid schemes with OCI and traditional SI \cite{hoagland_hybrid_2021}.
Both resynchronize cells by sweeping to improve convergence; however, the resulting algorithms are no longer space-parallel and involve a potentially more-expensive sweep operation.

\section{Conclusions and Future Work}

We derived the multiple balance and simple corner balance time-space discretization schemes and demonstrated, with Fourier analysis, that our time iteration method is unconditionally stable.
We also derived eigensystems for one-cell inversion and source iteration, showing that one-cell inversion iterations converge faster as mean free time shrinks.
Furthermore, OCI's convergence rate improves faster than SI's with decreasing mean free time.
We confirmed this with both Fourier and empirical analysis of implemented one-cell inversion and source iteration solvers.
Although we only explored block Jacobi OCI, we also expect this behavior to improve convergence of time-dependent block Gauss--Seidel OCI.

When more iterations are required to converge problems of interest---particularly in highly scattering and optically thin problems---OCI can run individual iterations significantly faster than SI when using batched direct solvers on GPUs from vendor-supplied libraries.
For OCI the number of on-device performant compute kernels is limited to data-parallel matrix-building operations, with all other compute kernels being called from optimized libraries. 
While optimization could improve both the OCI and SI algorithms, we analyzed performance to ensure there was little computational overhead from data movement and user-defined kernels.

Moving forward, we are exploring synthetic acceleration techniques to preserve the OCI space-parallel performance on GPUs while ameliorating issues in the thin and scattering limits.
Space-parallel OCI schemes offer promise as a high-performing class of iterative solvers for time-dependent radiation transport on modern heterogeneous compute architectures.

\section*{Acknowledgments}

The authors thank Dmitriy Anistratov of North Carolina State University for useful conversations about Fourier analysis results, James Warsa of Los Alamos National Laboratory for useful conversations about previous work and Damon McDougall of Advanced Micro Devices for support using ROCm compilers and profilers. 
The authors also thank the high performance computing staff at Lawrence Livermore National Laboratory for support using the Tioga machine.

This work was supported by the Center for Exascale Monte Carlo Neutron Transport (CEMeNT), a PSAAP-III project funded by the Department of Energy, grant number: DE-NA003967.

\bibliographystyle{IEEEtran}
\bibliography{main}

\appendix
\section{Source Iteration Systems}
\label{app:source_iteration}

\newcommand{\lph}{^{(l+1/2)}}
\newcommand{\lp}{^{(l+1)}}

Using a source iteration to solve a multi-group problem with multiple balance in time \cite{variansyah_robust_2021} and simple corner balance in space \cite{adams_subcell_1997} gives a $4\times 4$ system of linear equations:
\begin{equation}
    \label{eq:app:si_sys}
    \mathbf{L}_{c,j,g,m} \begin{bmatrix}
    \psi_{m,k,g,j,L}\lph \\
    \psi_{m,k,g,j,R}\lph \\
    \psi_{m,k+1/2,g,j,L}\lph \\
    \psi_{m,k+1/2,g,j,R}\lph
    \end{bmatrix}
    = \mathbf{d}_{j,m,g} \;,
\end{equation}
solved in every angle and group by sweeping from cell to cell,
where $\mathbf{L}_{c,j,g,m}$ is from Eq. \eqref{eq:Aj}
and
\begin{subequations}
\label{eq:app:si_rhs}
\begin{equation}
    \mathbf{d}_{j,g,m} = 
    \begin{cases}
        \bm{d}_{j,g,m}^+ & \mu_m>0 \\
        \bm{d}_{j,g,m}^- & \mu_m<0 \\
    \end{cases} \;,
\end{equation}
where
\begin{equation}
   \bm{d}_{j,g,m}^+ = \begin{bmatrix}
    \frac{\Delta x_j}{4} \left( \Sigma_{s,g\rightarrow g,j} \phi_{k,g,j,L}^{(l)}  +   Q_{k,j,L} \right) + \frac{\Delta x_j}{2 v_g \Delta t} \psi_{m,k-1/2,g,j,L} + \mu_m \psi_{m,k,g,j-1,R}\lph \\
    \frac{\Delta x_j}{4} \left( \Sigma_{s,g\rightarrow g,j} \phi_{k,g,j,R}^{(l)}  +   Q_{k,j,R} \right) + \frac{\Delta x_j}{2 v_g \Delta t} \psi_{m,k-1/2,g,j,R} \\
    \frac{\Delta x_j}{4} \left( \Sigma_{s,g\rightarrow g,j} \phi_{k+1/2,g,j,L}^{(l)}  +   Q_{k+1/2,j,L} \right) + \mu_m \psi_{m,k+1/2,g,j-1,R}\lph \\
    \frac{\Delta x_j}{4} \left( \Sigma_{s,g\rightarrow g,j} \phi_{k+1/2,g,j,R}^{(l)}  +   Q_{k+1/2,j,R} \right) 
    \end{bmatrix} \;,
\end{equation}
\begin{equation}
   \bm{d}_{j,g,m}^- = \begin{bmatrix}
    \frac{\Delta x_j}{4} \left( \Sigma_{s,g\rightarrow g,j} \phi_{k,g,j,L}^{(l)}  +   Q_{k,j,L} \right) + \frac{\Delta x_j}{2 v_g \Delta t} \psi_{m,k-1/2,g,j,L}  \\
    \frac{\Delta x_j}{4} \left( \Sigma_{s,g\rightarrow g,j} \phi_{k,g,j,R}^{(l)}  +   Q_{k,j,R} \right) + \frac{\Delta x_j}{2 v_g \Delta t} \psi_{m,k-1/2,g,j,R} - \mu_m \psi_{m,k,g,j+1,L}\lph  \\
    \frac{\Delta x_j}{4} \left( \Sigma_{s,g\rightarrow g,j} \phi_{k+1/2,g,j,L}^{(l)}  +   Q_{k+1/2,j,L} \right)  \\
    \frac{\Delta x_j}{4} \left( \Sigma_{s,g\rightarrow g,j} \phi_{k+1/2,g,j,R}^{(l)}  +   Q_{k+1/2,j,R} \right) - \mu_m \psi_{m,k+1/2,g,j+1,L}\lph
    \end{bmatrix} \;,
\end{equation}
and $\phi$ is the scalar flux.
\end{subequations}
After sweeping the mesh cells in the appropriate directions for each angle in the quadrature set and every group, the scalar flux vector can be updated via
\begin{equation}
\label{eq:app:si_sf}
\bm{\phi}_{k,g,j}\lph = 
\begin{bmatrix}
    \phi_{k,g,j,L}\lph \\
    \phi_{k,g,j,R}\lph \\
    \phi_{k+1/2,g,j,L}\lph \\
    \phi_{k+1/2,g,j,R}\lph
    \end{bmatrix}   = \sum\limits_{n=1}^N w_n 
    \begin{bmatrix}
    \psi_{n,k,g,j,L}\lph \\
    \psi_{n,k,g,j,R}\lph \\
    \psi_{n,k+1/2,g,j,L}\lph \\
    \psi_{n,k+1/2,g,j,R}\lph
    \end{bmatrix} \;.
\end{equation}
Then, group-to-group communication can be computed with
\begin{equation}
    \label{eq:app:si_update}
    \bm{\phi}_g\lp = \bm{\phi}_g\lph + \sum_{g'\neq g} \Sigma_{s, g'\rightarrow g } \bm{\phi}_{g'}\lph \; .
\end{equation}
Note that within group scattering is computed in the transport sweep itself, in Eq.~\eqref{eq:app:si_rhs}.
This algorithm allows for all groups and angles to be solved in parallel (using a Jacobi iteration). 
This algorithm gives source iterations the greatest number of degrees of freedom to parallelize for a 1D, slab geometry, multi-group problem---the best case scenario for GPU performance.
\begin{subequations}
Then, the source iteration can continue until 
\begin{equation}
    ||\bm{\phi}^{(l+1)}-\bm{\phi}^{(l)}||_{2} < \epsilon(1-\rho_{e,SI}) \; ,
\end{equation}
where $\epsilon$ is the convergence tolerance and
\begin{equation}
    \rho_{e,SI} = \frac{||\bm{\phi}^{(l+1)}-\bm{\phi}^{(l)}||_{2}}{||\bm{\phi}^{(l)}-\bm{\phi}^{(l-1)}||_{2}} \; ,
\end{equation}
\end{subequations}
is an empirical estimation of the spectral radius computed at every iteration of a transport solve. 
After converging, the simulation can move to the next time step.

\end{document}